\documentclass[apsx,12pt,tightenlines,amsmath,amssymb]{revtex4x}
\usepackage{graphicx}
\usepackage{dcolumn}
\usepackage{bm}
\usepackage{epsfig}

\def\Eqref#1{Eq.~(\ref{#1})}
\newcommand{\be}{\begin{equation}}\newcommand{\ee}{\end{equation}}
\newcommand{\bea}{\begin{eqnarray}}\newcommand{\eea}{\end{eqnarray}}
\def\Tr{\mathop{\mathrm{Tr}}\nolimits}

\def\evU{^{(U)}}
\def\beff{\beta_a\evU}

\def\AplusB{A$\mskip-2mu+\mskip2mu$B}

\begin{document}

\title{Dissipation, interaction and relative entropy}

\author{B. Gaveau}
\affiliation{Laboratoire analyse et physique math\'ematique, 14 avenue F\'elix Faure, 75015 Paris, France}

\author{L. Granger}
\affiliation{Max Planck Institute for the Physics of Complex Systems, N\"othnitzer Str.\ 38, D-01187 Dresden, Germany}

\author{M. Moreau}
\affiliation{Universit\'e Pierre et Marie Curie, LPTMC, case 121, 4 pl.\ Jussieu, 75252 Paris cedex 05, France}

\author{L. S. Schulman}
\affiliation{Physics Department, Clarkson University, Potsdam, New York 13699-5820, USA}
\email{schulman@clarkson.edu}

\date{\today}
\begin{abstract}
Many thermodynamic relations involve inequalities, with equality if a process does not involve dissipation. In this article we provide equalities in which the dissipative contribution is shown to involve the relative entropy (a.k.a. Kullback-Leibler divergence). The processes considered are general time evolutions both in classical and quantum mechanics, and the initial state is sometimes thermal, sometimes partially so. As an application, the relative entropy is related to transport coefficients.
\end{abstract}

\maketitle

\section{Introduction\label{s.intro}}

Dissipation reflects the profound distinction between work and heat in finite-time dynamical processes. In this article we relate dissipation to \textit{relative entropy} a quantity that we use to distinguish between actual and idealized time evolution as we explain below.

In thermodynamics and in kinetic theories the true state of a system is replaced by an idealized coarse-graining and as a consequence, the true evolution is replaced by an idealized evolution of the corresponding coarse-grained state (or by a quasi-static evolution in thermodynamics). There are two good reasons for using this idealization:\\
\indent1. It is impossible---even in principle---to specify the exact state of a large system. An attempt at extremely high precision would modify the system, even in a classical context (related to Maxwell's demon). And it is even worse for quantum systems.
\\
\indent2. Only slow variables can be measured with confidence and stability. As a result, an observer can only describe the system as a state of minimal information (or maximal entropy) compatible with the observed slow variables \cite{jaynes1, jaynes2, landau}. During the idealized evolution, information is lost and the entropy of the idealized state increases; however, under symplectic or unitary evolution the entropy of the true state remains constant.

The difference between the real and the idealized evolutions is measured by the dissipation, both for the information content and for the slow variables, in particular for energy. Entropy increase, for example, is a measure of loss of information concerning the actual processes compared to the coarse-grained evolution.

Standard thermodynamics uses the maximal coarse graining of equilibrium, and the idealized evolution is not modeled explicitly, so the dissipation is taken into account only by inequalities. For more detailed coarse-graining (as in hydrodynamics, Boltzmann's equation, kinetic theories or stochastic thermodynamics) one can obtain an estimate for the dissipative effects, for example, by the calculation of transport coefficients.

In this article, instead of choosing a specific model of coarse-grained evolution, we use exact dynamics and the fact that entropy remains constant during evolution. At that level of generality our results are model-independent identities. Dissipation comes from the loss of average microscopic information between the exact state and the information carried by the initial state, but averaged over the actual state: this quantity is exactly the relative entropy (also known as the Kullback-Leibler divergence \cite{cover}) between the actual true state and the coarse-grained initial state. As this quantity is always positive, it provides a lower bound for the interaction energy or an upper bound for the work that one can extract from the system. Moreover, these bounds are attained if and only if the state after evolution is the same as the initial state, in which case no work is extracted from the system.

The many equalities and inequalities given below are often more precise versions of known relations. In each expression that we provide it is precisely in the relative entropy of the various initial and final states that the dissipation lies, justifying the assertions made above.	

In the following material, we first consider a two-component system, A and B. When A begins in thermal equilibrium we find an identity for the energy transfer to A to achieve a given reduction of the information content of B. This identity contains explicit relative entropy terms. The Brillouin-Landauer inequality is an immediate consequence. Following that, we take initial thermal states for both A and B and obtain a Clausius identity as well as an identity for the variation of the interaction energy before and after the evolution. We next introduce an external agent which varies a control parameter for A. This leads to an identity, again involving relative entropy, for the work performed on the whole system, \AplusB, in terms of the difference of free energies before and after the evolution. Continuing, we study the effect of an external agent on an (otherwise) isolated system; again the work is given in terms of an identity relating the difference of internal energies and the usual dissipative terms. Finally, we show, in a quantum context, that the heat conductivity between two systems in interaction can be obtained from the relative entropy, confirming that relative entropy terms do measure the dissipation. In some of our examples one or both systems are initially at thermal equilibrium, but only the initial temperatures appear explicitly and no coarse graining by an effective final or intermediate thermal state is used.

\section{Definitions and notation.\label{s.definitions}}

Denote the state of a classical or quantum system by $\rho$\@. ``$\Tr A$'' indicates either an integral in phase space (classical observable) or the trace in Hilbert space (quantum observable). For both, $\rho\ge0$ and $\Tr\rho=1$\@. The entropy of $\rho$ is defined as $S(\rho)\equiv-\Tr\rho\log\rho$\@. For states $\rho$ and $\rho'$, the relative entropy is
$S(\rho|\rho')\equiv\Tr\left[\rho\left(\log\rho-\log\rho'\right)\right]$\@. It is known \cite{cover} that $S(\rho|\rho')\ge0$, with equality iff $\rho=\rho'$\@. It is identically true that
\be
S(\rho|\rho')=S(\rho')-S(\rho)-\Tr\left[\left(\rho-\rho'\right)\log\rho'\right]
\,,
\label{e.bg1}
\ee
and this identity lies behind all results in the present article. If $U$ is an evolution operator for some time interval, either symplectic or unitary, and $A$ an observable, let $A^{(U)}$ denote the evolute of $A\,$; e.g., in the quantum case, $\rho^{(U)}=U\rho U^\dagger$\@. For $\Phi$ a function of $\rho$, we also define $\delta\evU \Phi(\rho) \equiv \Phi(\rho\evU)-\Phi(\rho)$\@.

\section{Information and energy transfers in interacting systems.\label{s.twosystems}}

Let A and B denote interacting systems, with respective Hamiltonians $H_A$ and $H_B$, and interaction energy $V$\@. Energy expectations are $E_A(\rho)=\Tr(H_A\rho)$, $E_V(\rho)=\Tr(V\rho)$, etc. Energy conservation under evolution by $U$ requires
\be
\delta\evU\left[E_A(\rho)+ E_B(\rho) + E_V(\rho)\right]=0
\label{e.bg2}
\,.
\ee
For $\rho$ a state of \AplusB, let $\rho_A\equiv\Tr_B\rho$ and $\rho_B\equiv\Tr_A\rho$\@. We take the initial (time-0) state of \AplusB\ to be
\be
\rho_0=\rho_A(\beta_A)\otimes\rho_{B0}
\label{e.bg3}
\,,
\ee
where $\rho_A(\beta_A)=\exp(-\beta_A H_A)/Z_A(\beta_A)$ is the thermal state at temperature $T_A\equiv\beta_A^{-1}$ and $\rho_{B0}$ is any state of B\@. We immediately deduce from \Eqref{e.bg1}
\be
\beta_A \delta\evU E_A(\rho)=-\delta\evU S(\rho_B)+
         \left[ S(\rho\evU|\rho_A\evU\otimes\rho_B\evU)+ S(\rho_A\evU|\rho_A(\beta_A))
         \right]
\label{e.bg4}
\,.
\ee
In particular, if the evolution $U$ is such that $\delta\evU S(\rho_B)\le0$ (so that the information content of B has increased), then
\be
\beta_A \delta\evU E_A(\rho)\ge |\delta\evU S(\rho_B)|
\label{e.bg5}
\,.
\ee
This implies that the energy of A has \textit{increased} (or that the energy $E_B(\rho) + E_V(\rho)$ has decreased). Thus a transfer of information to B implies a transfer of energy from B to A \cite{brillouin2, landauerIBM1961, landauerBerichte1976, landauerPhysLett1996, landauerPhysicaA1993}. Moreover, the equality (\ref{e.bg5}) is attained only when $\rho_A\evU=\rho_A(\beta_A)$ and $\rho\evU = \rho_A\evU\otimes\rho_B\evU$, in which case $\delta\evU E_A(\rho)=0$, $\delta\evU S(\rho_B)=0$, and no transfer has occurred.
Finally, if $\rho_A\evU=\rho_A(\beta_A)$, then $\delta\evU E_A(\rho)=0$ and from \Eqref{e.bg5}, $\delta\evU S(\rho_B)=0$\@. In particular, if A is a true thermal bath (unchanged by the evolution) it cannot be used to lower the entropy of another system, B\@.

\smallskip

\noindent\textsf{Remark}:
In \Eqref{e.bg4} the expression in square brackets is positive. Note too that $S(\rho\evU|\rho_A\evU\otimes\rho_B\evU)$ is the mutual information between the A and B in the state $\rho_A\evU$.

\section{When the systems are initially in equilibrium.\label{s.equil}}

Assume now that A and B are in thermal equilibrium states (with $T_B\equiv\beta_B^{-1}$) at $t=0$ so that
\be
\rho_0=\rho_A(\beta_A)\otimes\rho_B(\beta_B)
\label{e.bg6}
\,.
\ee
Then the following identities can be deduced
\be
-\delta\evU E_V(\rho)=\left(1-\beta_A/\beta_B\right)
                             \delta\evU E_A(\rho) +T_B S(\rho\evU|\rho_0)
\label{e.bg7}
\,,
\ee
\be
\beta_A \delta\evU E_A(\rho)+\beta_B \delta\evU E_B(\rho)=S(\rho\evU|\rho_0)
\label{e.bg8}
\,,
\ee
\bea
\delta\evU E_A(\rho)+ \delta\evU E_B(\rho)&=&
     T_A \delta\evU S(\rho_A) + T_B \delta\evU S(\rho_B)+
     \nonumber\\
     &&+\left[
       T_A \delta\evU S(\rho_A\evU|\rho_A(\beta_A))
      + T_B \delta\evU S(\rho_B\evU|\rho_B(\beta_B))
        \right]
\label{e.bg9}
\,.
\eea
From this one can derive corresponding inequalities, for example
\be
-\delta\evU E_V(\rho)\ge\left(1-\beta_A/\beta_B\right)\delta\evU E_A(\rho)
\label{e.bg7prime}
\,,
\ee
\be
\beta_A \delta\evU E_A(\rho)+\beta_B \delta\evU E_B(\rho)\ge0
\label{e.bg8prime}
\,,
\ee
\be
\delta\evU E_A(\rho)+ \delta\evU E_B(\rho)\ge
     T_A \delta\evU S(\rho_A) + T_B \delta\evU S(\rho_B)
\label{e.bg9prime}
\,.
\ee
These relations imply important theoretical and practical conclusions. If, for instance, $T_A > T_B$  and $\delta\evU E_A(\rho)<0$, it follows from (\ref{e.bg7prime}) that $\delta\evU E_V(\rho)<0$ and that the interaction energy cannot be neglected, contrary to current approximations. On the other hand, one sees that (\ref{e.bg8prime}) is the Clausius inequality. It should be pointed out that the equalities (\ref{e.bg7}-\ref{e.bg9}) are much stronger than the corresponding inequalities (\ref{e.bg7prime}-\ref{e.bg9prime}). In particular, (\ref{e.bg7}-\ref{e.bg9}) show that these inequalities are changed into equalities iff there are no changes under the evolution operator $U$\@. On the other hand, the last term in the right hand side of (\ref{e.bg9}) can be interpreted as the energy dissipation, which is thus expressed in terms of relative entropies.

\section{Interacting systems coupled to a work source.\label{s.work}}

As before, A and B interact, but now A is coupled to an external source of work and its Hamiltonian becomes $H_A(\lambda)$, with the parameter $\lambda$ taking the value $\lambda_0$ at time-0. The initial state is now written $\rho_0=\rho_{A,0}\otimes\rho(\beta_B)$ (so B is in a thermal state, but A may not be). The external observer can modify $\lambda$, and at the end of the evolution $U$, $\lambda$ has some value $\lambda\evU$\@.
Then the work $\delta\evU W$ received by the external observer is
\be
\delta\evU W=-\delta\evU E_V(\rho)-\delta\evU F_A(\rho_A,\beta_B)
            -T_B\left[S(\rho\evU|\rho_A\evU\otimes\rho_B\evU + S(\rho_B\evU|\rho_B(\beta_B))\right]
\,,
\label{e.200}
\ee
where
\be
\delta\evU F_A(\rho_A,\beta_B)
         = \bigl(E_A(\rho_A\evU)-T_B S(\rho_A\evU)\bigr)
                - \bigl(     E_A(\rho_{A,0})-T_B S(\rho_{A,0}) \bigr)
\,,
\label{e.210}
\ee
is the variation of the (non-equilibrium) free energy of A, calculated at temperature $T_B$, and
\be
E_A(\rho_A\evU)=\Tr\left(\rho_A\evU H_A(\lambda\evU)\right)
\,,\quad E_A(\rho_{A,0})=\Tr\left(\rho_{A,0} H_A(\lambda_0)\right)
\,.
\label{e.220}
\ee
In particular we deduce
\be
\delta\evU W\le-\delta\evU E_V(\rho)-\delta\evU F_A(\rho_A,\beta_B)
\,.
\label{e.230}
\ee
This inequality is analogous to, but more general than, the standard thermodynamic inequality concerning the work that can be extracted isothermally. Here the process need not be isothermal, B is not a heat bath, and A's final state need not be equilibrium at temperature $T_B$\@. Moreover, we have equality in \Eqref{e.230} iff $\rho_B\evU=\rho_B(\beta_B)$ and $\rho\evU=\rho_A\evU\otimes\rho_B(\beta_B)$\@. Assume now that A and B are initially in thermal states at $T_A$ and $T_B$, so that $\rho_0=\rho_A(\beta_A,\lambda_0)\otimes\rho_B(\beta_B)$. The work \textit{received} by the external observer is then
\bea
\delta\evU W&=&
-\delta\evU E_V(\rho) -\left(1-\beta_B/\beta_A\right)\delta\evU E_B(\rho)
   +\left(F_A(\beta_A,\lambda_0)-F_A\evU\right)
 \nonumber\\
&&\qquad -T_A
  \left[
    S(\rho\evU|\rho_A\evU\otimes\rho_B\evU) +S(\rho_B\evU|\rho_B(\beta_B))
  \right]
\label{e.bg10}
\,,
\eea
where $F_A(\beta_A,\lambda_0)$ is the equilibrium free energy of A and $F_A\evU$ is the final (non-equilibrium) free energy at temperature $T_A$,
\be
F_A\evU=\Tr \left(H_A(\lambda\evU)\rho_A\evU\right)-T_AS(\rho_A\evU)
\label{e.bg11}
\,.
\ee
If $\beta_A=\beta_B$ one obtains the inequality
\be
\delta\evU W\le -\delta\evU E_V(\rho) + F_A(\beta_A,\lambda_0)-F_A\evU
\label{e.bg12}
\,,
\ee
with equality iff $\rho_B\evU=\rho_B(\beta_B)$ and $\rho\evU=\rho_A\evU\otimes \rho_B(\beta_A)$\@. \Eqref{e.bg12} is more general than the usual thermodynamic relation \cite{landau} in two ways: it includes the interaction energy $\delta\evU E_V(\rho)$ and it involves a non-equilibrium free energy $F_A\evU$ (defined by \Eqref{e.bg11}) which, being equal to or greater than the equilibrium value, strengthens the inequality.

\section{Single system coupled to a work source.\label{s.single}}

A single system has Hamiltonian $H(\lambda)$, with changes in the external parameter $\lambda$ corresponding to work. At $t=0$, $\lambda=\lambda_0$, and we assume the state to be thermal: $\rho_0=\exp(-\beta_0 H(\lambda_0))/Z(\beta_0,\lambda_0)$\@. The external observer changes $\lambda$ to $\lambda\evU$ inducing an evolution $U$ of the system which ultimately reaches the state $\rho\evU$\@. Define the  \textit{adiabatic} temperature $\beff$ by
\be
S(\rho\evU)=S(\beff,\lambda\evU)
\label{e.bg13}
\,,
\ee
with $S(\beta,\lambda)$ the entropy of the thermal state \cite{note:effective}. Then one has
\be
\delta\evU W=E(\beta_0,\lambda_0)-E(\beff,\lambda\evU) - \frac1{\beff}S(\rho\evU|\rho(\beff,\lambda\evU))
\label{e.bg14}
\,,
\ee
with $E(\beta,\lambda)$ the energy of the thermal state. From this one deduces the standard \cite{landau} thermodynamic inequality for adiabatic processes
\be
\delta\evU W\le E(\beta_0,\lambda_0)-E(\beff,\lambda\evU)
\label{e.bg14prime}
\,,
\ee
with equality iff $\rho\evU=\rho(\beff,\lambda\evU)$\@. In particular, if the external observer imposes a cycle, namely $\lambda\evU=\lambda_0$, then $\beff=\beta_0$ and
\be
\delta\evU W =-\frac1{\beta_0}S(\rho\evU|\rho(\beta_0,\lambda_0))
\le 0
\label{e.bg15}
\,.
\ee

\smallskip

\noindent\textsf{Remark}:
The inequality \Eqref{e.bg14prime} is different from Jarzynski's inequality \cite{jarzynski} because \Eqref{e.bg14prime} contains the adiabatic temperature, rather than the initial temperature, $\beta_A$\@.

\section{Relative entropy and transport coefficients.\label{s.transport}}

The previous identities can be transformed into inequalities similar to the standard inequalities of non-equilibrium thermodynamics using the positivity of the relative entropies appearing in these identities. Thus the relative entropy terms measure exactly the dissipative effects, coming from the fact that actual evolutions differ from idealized quasi-static processes. As an example, we deduce the transport coefficients, confirming the significance of the relative entropy.

Consider systems A and B, initially in equilibrium at temperatures $\beta_A^{-1}$ and $\beta_B^{-1}$, respectively. The joint system \AplusB\ evolves until a final time $t$\@. Let $U$ or $U(t)$ denote the overall evolution operator from 0 to $t$. We wish to evaluate the flow of energy from A to B, and deduce an expression for the thermal conductivity.

Let $\rho(t)$ be the state at time-$t$\@. Then as shown above (\Eqref{e.bg7})
\be
\left(\beta_A-\beta_B\right)
   \delta\evU E_A(\rho_A)
   =\beta_B\delta\evU E_V(\rho)
                           +S(\rho(t)|\rho_A(\beta_A)\otimes\rho_B(\beta_B))
\label{e.bg16}
\,.
\ee
We can estimate each term of the second member of \Eqref{e.bg16} in the Born approximation, for systems A, B in the limits of continuous spectrum and $t$ large. Then the interaction term $\delta\evU E_V(\rho)$ becomes negligible and in a calculation whose details we will present elsewhere \cite{gaveaufuture} one obtains
\be
   \delta\evU E_A(\rho_A)
   \approx
    \frac{S(\rho(t)|\rho_A(\beta_A)\otimes\rho_B(\beta_B))}{\beta_A-\beta_B}
    \approx
    \left(\beta_A-\beta_B\right)Kt
\label{e.bg17}
\,.
\ee
In this expression the term of principal interest is $K$, a thermal conductivity coefficient proportional to the square of the matrix element of the interaction energy $V$\@. In particular
\be
K=\frac\pi{\hbar Z_A(\beta_A) Z_B(\beta_B)}
  \int dE_A\, dE'_A\, dE_B\, dE'_B\, \phi(E_A, E'_A, E_B, E'_B)
\label{e.150}
\,,
\ee
where
\bea
\phi&=& f_A(E_A) f_A(E'_A) f_B(E_B) f_B(E'_B)e^{-\beta_A E_A'-\beta_B E_B'}|V(E_A',E_B'|E_A,E_B)|^2\times
\nonumber \\
&&\quad\times
\delta(E_A+E_B-E'_A- E'_B)\, \frac{E_A-E_A'}{\beta_A -\beta_B}
\left(
1-e^{(\beta_A -\beta_B)(E_A-E_A')}
\right) >0
\label{e.160}
\,,
\eea
and $f_A$ and $f_B$ are the densities of states of systems A and B\@. \Eqref{e.bg17} is a form of the Fourier heat law.

\section{Conclusions.\label{s.conclusions}}

Eqs.\ (\ref{e.bg4}) and (\ref{e.bg9}) are identities for the energy exchange or information exchange between parts of a classical or quantum system evolving under the exact true dynamics, starting from a thermal state. The state of the system during its evolution is the true state and is a non-equilibrium state. In particular its entropy remains constant during the evolution.  Thus there is no coarse-graining of the evolution and these identities do not depend on a choice of a specific model (stochastic or kinetic). These identities contain \textit{relative entropy} terms between the actual evolved state and the initial state, and because the relative entropy is positive, the consequences of these identities are inequalities which generalize the inequalities of non-equilibrium thermodynamics. See Eqs.\ (\ref{e.bg5}) and (\ref{e.bg7prime})-(\ref{e.bg9prime}). The relative entropy terms can be considered as general expressions measuring dissipation effects occurring internally between interacting parts of the system during its evolution. This is confirmed by the fact that one can deduce the thermal conductivity coefficient between two interacting parts of a system from an estimate of the relative entropy, as is seen in Eqs.\ (\ref{e.bg17}) and (\ref{e.150}). The same method also gives an identity, Eqs.\ (\ref{e.bg10}) and (\ref{e.bg14}), for the work exchanged between a system and an external agent and thus an inequality (Eqs.\ (\ref{e.bg12}) and(\ref{e.bg14prime})) for the work that can be extracted from a system, together with an exact expression for the dissipation. We have also seen that the interaction energy, which is often neglected, can be used to give additional precision (cf.\ \Eqref{e.bg10}) and is important in calculating transport coefficients (see Eqs\ (\ref{e.150}),~(\ref{e.160})).

\begin{acknowledgments}
We are grateful to the Max Planck Institute for the Physics of Complex Systems, Dresden, for its gracious hospitality.
\end{acknowledgments}

\section*{References} \vskip-50pt


\begin{thebibliography}{13}
\bibitem[{not()}]{note:effective}
\bibinfo{note}{Another choice is what one can call the \textit{effective}
  temperature, namely $\beta$ such that
  $E(\rho\evU)=E(\beta_e\evU,\lambda\evU)$, or more generally, given $E$,
  $\beta_e$ such that $E=\Tr \exp(-\beta_e H) )/Z(\beta_e)$\@.}
\end{thebibliography}
\end{document}